\documentclass[11pt]{article}

\usepackage{geometry}
\usepackage{graphicx}
\usepackage{url}

\title{SVEN: Informative Visual Representation of Complex Dynamic Structure %
\thanks{Distribution A: Approved for public release; distribution unlimited.
88ABW
Cleared 10/31/2014; 88ABW-2014-5108}
}

\author{Dustin L. Arendt%
\thanks{D. Arendt is currently a research scientist at Pacific Northwest National Laboratory.  email: dustin.arendt@pnnl.gov} \\
Air Force Research Laboratory \and %
Leslie M. Blaha%
\thanks{email: leslie.blaha@us.af.mil} \\
 Air Force Research Laboratory}

\usepackage{subfigure}
\usepackage{url}

\newcommand{\myfigure}{\begin{figure}[tbp]}

\begin{document}

\maketitle

\begin{abstract}
Graphs change over time, and typically variations on the small multiples or animation pattern is used to convey this dynamism visually.  However, both of these classical techniques have significant drawbacks, so a new approach, Storyline Visualization of Events on a Network (SVEN) is proposed.  SVEN builds on storyline techniques, conveying nodes as contiguous lines over time.  SVEN encodes time in a natural manner, along the horizontal axis, and optimizes the vertical placement of storylines to decrease clutter (line crossings, straightness, and bends) in the drawing. This paper demonstrates SVEN on several different flavors of real-world dynamic data, and outlines the remaining near-term future work.
\end{abstract}

\section{Introduction} 
Graphs have become an essential mathematical tool to aid in the representation and analysis of relationships between discrete entities such as people, companies, computers, etc.  Not surprisingly, visualization of graphs (i.e., graph drawing) has been an active area of research for the past several decades.  However, the world is a dynamic place, so graphs that model objects and their relationships in the world often change over time.  To account for this, one often chooses a particular time interval of interest and constructs a single graph capturing the relationships observed in that interval.  This is the {\it de facto} method for social network analysis, where the graph is constructed from a series of interviews, questionnaires, or observations occurring within some time interval \cite{wasserman1994}.  Visualization of the graph often plays a crucial role in the initial understanding of the phenomena of interest.

\begin{figure}
\begin{center}
\includegraphics[width=.35\textwidth]{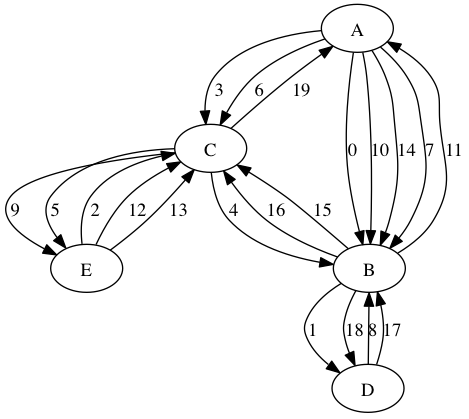}
\end{center}
\caption{Representation of a sequence of events on a network as a directed labeled multigraph.  Labels on edges represent the time of the event.}
\label{fig:multigraph}
\end{figure}

There is no reason to require that such data must be compressed into a single graph aside from convenience.  On the contrary, one would expect a {\it longitudinal} analysis to potentially yield more insight into dynamic phenomena than a more restrictive static analysis.  A typical way to understand a dynamic graph is to induce a sequence of subgraphs 
\[\{G_1=(V_1,E_1), G_2=(V_2,E_2), ...\},\]
by partitioning a time interval into several smaller non-overlapping time windows.  Alternatively we can consider a sequence of interaction events on a continuous timescale as a set of time-edge tuples of the form
\[\{(t_1,(u_1,v_1)), (t_2,(u_2,v_2)), ...\},\]
where the $i^{th}$ interaction occurs at time $t_i$ and involves $u_i$ and $v_i$, and reads as ``at time $t_i$ node $u$ interacted with node $v$.''  For simplicity we assume that interactions are pairwise and of negligible duration relative to the timescale of the visualization; this type of network has been referred to as a ``contact sequence''~\cite{holme2012temporal}, and an example is shown in Fig.~\ref{fig:multigraph}.  Many relevant real-world phenomena can be described in this manner, such as a group of authors publishing a paper together, money being wired between bank accounts, one or more proteins activating or inhibiting another protein, and dialogue between characters a play or novel.

Given that node-link visualizations of static graphs are wildly popular, it is not surprising that demand for effective dynamic graph visualization capability is increasing.  Perhaps what is surprising is that, to date, no general purpose framework for dynamic graph visualization has become widely accepted in the same way that force directed placement has been for static graph visualization.  Research in dynamic graph visualization began with showing the changes to the graph either as a sequence of still images (e.g., small multiples) for each $G_i$, or as an animation that interpolates between these drawings \cite{moody2005dynamic,bender2006art}.  While this approach to dynamic graph visualization problem inherits the challenges of optimizing the aesthetic properties of the static views of the graphs, it also introduces a new problem of how to change those views in a way that doesn't confuse or mislead the viewer.  It is hypothesized that the change between consecutive drawings should small to help preserve the user's ``mental map'' \cite{purchase2007important,archambault2011}.

However, even if the mental map is well preserved, both small multiples and animation have significant drawbacks.  The small multiples technique must decrease the size of the individual drawings in order to fit the entire collection on screen, and it can be tasking to trace nodes and their relationships as they jump from one view to the next.  With animated drawings, motion can distract the user when that motion is an artifact of the layout algorithm and not reflective of an important change to the graph structure. Also, the user will be burdened by seeking back and forth within the animation to understand the important changes to the graph.  While annotating an image is as trivial writing on a printed version, annotating an animation requires specialized software, and the animation should be rendered in a standard format.  Furthermore, animated visualizations prevent multiple viewers from simultaneously viewing different times within the animation.

The issues created by the ``classical'' approaches to dynamic network visualization can be partially addressed through effective interactions and algorithms, and much research has been devoted to such iterative improvements.  However, the visualization community is still awaiting broad acceptance of a ``workhorse visualization'' for dynamic graph visualization.  While not originally intended as general purpose solution for dynamic graph visualization, storyline visualization \cite{ogawa2010software,kim2010tracing,tanahashi2012design,liu2013storyflow} holds promise for this.  The contribution of this paper is the adaptation of storyline visualization techniques to dynamic network visualization--storyline visualization of events on a network (SVEN)--an effective visualization technique for both flavors of dynamic networks described above.

Consistent with previous storyline visualization techniques, SVEN draws nodes as contiguous lines, and time is encoded on the horizontal dimension.  Nodes that are interacting are drawn close to each other, and interactions are drawn as curved vertical lines that connect the interacting nodes at the appropriate time.  SVEN contains an optimization algorithm to improve the aesthetic quality of the drawing by re-arranging and straightening the storylines, and renders the storyline using a metro map styling.  What follows is a discussion of relevant literature related to dynamic graph visualization, a brief outline of the requirements and design of SVEN, and an overview SVEN's layout algorithm, followed by several examples of layouts generated by SVEN. Integration of SVEN into a visual analytics software system and evaluation of SVEN is left as future work.

\section{Related Work}
Dynamic graph visualization has its origin in static graph visualization, and many modern graph drawing algorithms have their basis in the original Kamada-Kawai \cite{kamada} and Fruchterman-Reingold \cite{fruchterman1991drawing} ``force-directed'' algorithms.  These algorithms define a model of a physical system from the graph whose energy can be measured and consequently minimized to produce an aesthetically pleasing drawing, ideally.  Popular, modern, freely available graph drawing packages include GraphViz \cite{ellson2002}, Gephi \cite{bastian2009}, and D$^3$ \cite{bostock2011,dwyer2009}.

Though not as mature as static graph visualization, the field of dynamic graph visualization is growing quickly.  As stated previously, the majority of dynamic graph drawing approaches use small multiples \cite{chi1999} or animation \cite{moody2005dynamic,bender2006art}, but few have effective overviews that portray all the changes to the graph at once.  Among techniques that do, an often repeated technique is to layer the 2-D graph drawings from each time slice into a third dimension \cite{brandes2003,erten2004}.  This approach shows connections between nodes during a specific time slice as well as the persistence of nodes over time slices, but the layering creates occlusion which decreases usability.  CiteSpace-II analyzes and visualizes trends in academic research over time \cite{chen2006}.  One of its visualizations overlays a node-link diagram on a timeline, with nodes representing events such as the publication of a scientific paper.

Network analysts can often still gain insight when provided only the ego networks (the subgraph containing the ego and its direct neighbors) of individuals instead of the entire network.  Based on this, Shi et al. developed a ``1.5-dimensional'' dynamic network visualization capable of showing the ego network of a particular individual of interest over time in a single picture \cite{shi2011dynamic}.  Burch et al. adapted the classical parallel coordinates visualization for dynamic networks \cite{burch2011parallel}.  In their system, vertices are drawn on parallel axes, the area between each axis represents a time interval, and edges connect vertices in adjacent axes.  The inevitable problem of having an overwhelming number of edge crossings for larger datasets is addressed by reducing the opacity of the lines drawn.

Recently, there has been increased interest in automated algorithms for ``storyline visualization,'' a technique where time is encoded in the horizontal dimension and characters are drawn as contiguous lines \cite{ogawa2010software,kim2010tracing,tanahashi2012design,liu2013storyflow} inspired by an XKCD web comic showing presence of characters in scenes throughout various popular films \cite{xkcd}.  When groups of characters interact, their storylines are closer within that time interval, relative to time intervals in which they are not interacting.  However, before detailing these works it is useful to provide a more historical context, as there are several older techniques that are similar.

\subsection{Predecessors to Storyline Techniques}
To the author's knowledge, the earliest portrayal of events on a network with a spatial encoding of time are ``sequence diagrams,'' developed in the 1970s, which are used to understand timing and synchronization in distributed systems \cite{lamport1978time}.  These visualizations were originally hand drawn; nodes were represented as vertical parallel lines with time moving from the bottom to the top of the page.  Communication events between processes were represented as wavy lines connecting connecting the processes sending and receiving the message at the local times the message was sent and received.  Sequence diagrams are effective for understanding trivially small networks, but become difficult to scale up as more nodes are added.  The ordering of nodes in the diagram can be chosen to minimize crossings, but this is the Traveling Salesman Problem, and even if an optimal solution was found, there is no guarantee that the resulting diagram would be interpretable for large or dense networks.

Visibility representations, which have existed since the mid 1980s, are also visually similar to storylines \cite{tamassia1986visibility}.  These are representations of planar graphs where nodes are drawn as horizontal lines, and edges are drawn as vertical lines connecting their endpoints.  Node-edge crossings are not allowed, so non-planar graphs do not have visibility representations. Due to this constraint, it is not apparent how visibility representations can be used as a general purpose solution for the dynamic graph visualization problem.  Nor is it apparent how to overload the horizontal axis in a visibility representation to also encode time.

Dwyer and Eades investigated using ``columns and worms'' to visualize dynamic network induced by movements of fund managers over time \cite{dwyer2002visualising}.  This is a 3-D visualization where nodes are represented as contiguous elements in the time dimension, and re-arranged in two spatial dimensions to optimize force-based graph aesthetics.  A disadvantage of this technique is that perspective distortion and occlusion can make it very difficult to accurately interpret timing and ordering of events.

Some researchers have simplified the dynamic graph visualization problem by increasing the level of abstraction and portraying how the network communities change over time \cite{rosvall2010mapping,reda2011visualizing}.  Visualizing this information is inherently simpler than visualizing the underlying dynamic network.  Rosvall and Bergstrom apply a significance clustering technique to the dynamic network at fixed time windows to partition the vertices into groups.  Then the flow of nodes between clusters at consecutive time windows is visualized using a technique similar to Sankey diagrams \cite{schmidt2008sankey} or parallel sets \cite{bendix2005parallel}, which they call ``alluvial diagrams.''  However, the authors make no attempt to improve the aesthetic quality of their visualizations by re-arranging nodes to reduce clutter (crossings between time windows). Instead nodes are ordered within each time window according to cluster size, which causes the diagrams to scale poorly as a function of the number of communities and time windows.

Another disadvantage of Rosvall and Bergstrom's technique is that significance clustering is performed on time windows independently, which might introduce noise (nodes oscillating between clusters over time), adding clutter and artifacts to the visualization. Berger-Wolf and Saia introduced an optimization based approach for dynamic community detection that overcomes this pitfall \cite{berger2006framework}, and this technique was improved in \cite{tantipathananandh2007framework,tantipathananandh2009constant}.  This framework was utilized to visualize dynamic communities for several datasets.  Communities, which span time, are represented as stacked rectangles that span the horizontal space of the visualization.  Similar to storyline visualizations, nodes are represented as horizontal lines, placed inside communities, and can only exist in one community at a time.  When a node changes communities a diagonal line is used to represent this change.  Edges in the network are not shown, and no attempt is made to optimize the ordering of the communities or the nodes within communities.  This would likely reduce clutter and improve the aesthetic properties of the visualization.

\subsection{Storyline Techniques}
Several recent visualizations \cite{kim2010tracing,ogawa2010software,tanahashi2012design,liu2013storyflow} appear to have been directly inspired by a series of hand drawn XKCD web comics \cite{xkcd} detailing character interactions in several popular films.  These visualizations all have a very similar visual appearance, and will be referred to as ``Storyline Visualizations'' in this paper. Kim et al. use storylines to simultaneously portray birth, death, marriage, and divorce within a genealogical diagram \cite{kim2010tracing}.  Their layout method is effective at producing an aesthetically pleasing drawing, but makes structural assumptions about the genealogical graph, preventing this technique from being applied to the general dynamic graph drawing problem.  Concurrently, Ogawa and Ma developed ``software evolution storylines,'' intended to be more generally applicable to dynamic graph drawing \cite{ogawa2010software}, which was later improved by Tanahashi and Ma \cite{tanahashi2012design}.  Due to the challenging multi-objective nature of the problem, their approach uses an evolutionary algorithm to simultaneously improve line wiggles, line crossovers, and whitespace gaps.

However, evolutionary algorithms are often computationally intensive, and when alternative solutions exist, they are frequently desirable.  For this reason, Liu et al. improved on previous work by developing a multi-stage hybrid optimization technique with essentially the same problem domain and aesthetic criteria \cite{liu2013storyflow}. Their contributions were algorithmic complexity improvements and the inclusion of hierarchical constraints on the interactions to improve the correctness of the drawings.  These latter two storyline visualization approaches both assume the data can be described in terms of ``interaction sessions,'' which are essentially data about the time, duration, and participants in an interaction event.  However, it is assumed that an individual cannot be in more than one interaction session at any given time, which continues to impose a certain structure on the data being visualized.  Thus, none of the storyline visualization techniques to date can be considered solutions for general purpose dynamic graph visualization.

\section{Requirements \& Design}
Dynamic graph visualizations based on small multiples or animation optimize aesthetics, but do not encode time spatially. Conversely, many of the dynamic graph visualization techniques (besides storyline visualizations) that do encode time spatially do not effectively optimize aesthetics.  Therefore, we propose two fundamental requirements for effective dynamic graph visualization:
\begin{itemize}
	\item Use space to represent time ({\bf R1}), and
	\item Reduce clutter ({\bf R2}) by minimizing the following
		\begin{itemize}
			\item node and edge crossings ({\bf R2a})
			\item node bends ({\bf R2b})
			\item edge length ({\bf R2c})
		\end{itemize}
\end{itemize}

We make a spatial encoding of time the first and most important requirement for effective dynamic graph visualization.  Encoding time spatially has one of the longest lasting precedents in the entire field of information visualization--the technique has been used for hundreds of years and many historical hand drawn examples exist \cite{aigner2011timebook}.  The second requirement inherits directly from the established visual aesthetics from classical node-link drawing techniques, and states that effective dynamic graph visualization should contain an algorithm that rearranges visual elements to decrease line crossings, line bends, and line length.  Below we describe how SVEN is designed with these requirements in mind.

Javed and Elmqvist compiled a useful set of common design patterns for composite visualizations \cite{javed2012exploring} that we find helpful--they allow us to describe how SVEN is constructed from simple visualization patterns.  In the language of their framework, SVEN consists of juxtaposed arc diagrams (one per time window) with integrated nodes.  Arc diagrams are a well known graph drawing technique where nodes are ordered and drawn in one dimension, and links are drawn as arcs of varying radius depending on the distance between nodes \cite{wattenberg2002arc}. Juxtaposition describes the placement of two or more views side-by-side, and integration is the practice of drawing lines to connect related objects in separate views.  In our case, when one node is present in two adjacent (juxtaposed) arc diagrams, they are connected by a line segment.  When detailed time information about events within a time window is known, the arcs within the arc diagram are shifted to the left or right accordingly.

\begin{figure}
\begin{center}
\includegraphics[width=.5\textwidth]{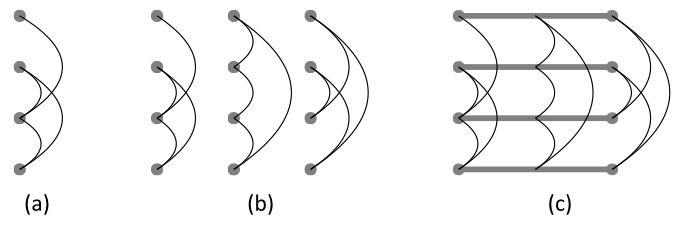}
\end{center}
\caption{(a) A single arc diagram (b) Juxtaposed arc diagrams (c) Integrated \& juxtaposed arc diagrams (SVEN)}
\label{fig:arc-diagrams}
\end{figure}

\begin{figure}
\begin{center}
\includegraphics[width=.5\textwidth]{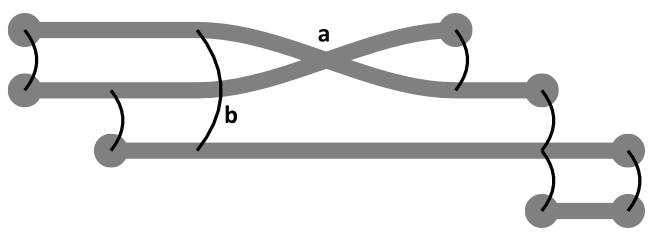}
\end{center}
\caption{An simple SVEN rendering with two time windows that contains one node-node crossing (a) and one node-edge crossing (b).}
\label{fig:cross}
\end{figure}

An example of constructing a SVEN drawing by juxtaposing arc diagrams is shown in Fig.~\ref{fig:arc-diagrams}.  For clarity, the four nodes are assigned the same four levels in each arc diagram.  However, this is not always the case, which may cause the integrating lines to cross, or to be drawn diagonally (instead of horizontally).  In SVEN, the integrating lines are always horizontal within a time window; crossing and diagonal lines occur only between time windows. We refer to the crossing of integration lines between time windows as ``node-node crossings.''  Additionally, edges (arcs in the time windows) may be crossed by nodes (lines integrating adjacent time windows), and we refer to this as ``node-edge crossings.''  A third type of crossing occurs between edges within each time widow are ``edge-edge crossings,'' but at the present time SVEN does not directly minimize edge-edge crossings (Fig.~\ref{fig:arc-diagrams} contains several edge-edge crossings).  Fig.~\ref{fig:cross} shows a idealized SVEN drawing that contains one node-node and one node-edge crossing.

The ordering, juxtaposition, and integration of arc diagrams for each time window directly supports {\bf R1}, by providing a spatial representation of time in the composite visualization.  Furthermore, the use of arc diagrams (instead of node-link diagrams, for example) makes more effective use of screen space by constructing a two-dimensional drawing out of a sequence of one dimensional drawings.  The ordering of each node within each time window can be adjusted to minimize node-node and node-edge crossings, supporting {\bf R2a} and {\bf R2c}.  The spacing between nodes within each time window can be adjusted to straighten the integrating lines, supporting {\bf R2b}.

Past storyline visualization techniques \cite{kim2010tracing,ogawa2010software,tanahashi2012design,liu2013storyflow} require that storylines move closer while the participants are interacting and then move apart afterwards.  We do not make this an explicit requirement in SVEN because interaction events are visually mapped onto arcs in the diagram, and for very complex interactions it may impossible or ambiguous to encode interaction as proximity.  Rather, proximity is more of a secondary visual cue in SVEN, and is minimized indirectly during the determination of node order within each time window.  We also note that many two-dimensional static graph drawing algorithms make the same tradeoff via force directed energy minimization schemes.  Few algorithms explicitly explicitly solve the layout based on the constraint that the nearest neighbors of a node in the graph drawing are also its neighbors in the graph, with Ref.~\cite{shaw2009structure} being a rare exception.  Finally, we note that omitting this requirement simplifies SVEN's layout algorithm, which we describe in detail in the following section.

\section{Layout Algorithm}
SVEN has five basic steps to transform raw data into a dynamic graph drawing:
\begin{enumerate}
\item {\bf Ingest:} Raw data is transformed into an graph sequence or event sequence partitioned into time windows, and from this an aggregate graph is constructed.
\item {\bf Order:} The ordering of nodes within each time window is chosen to minimize line crossings and length.
\item {\bf Align:} As many nodes as possible are chosen for alignment to minimize line bends.
\item {\bf Place:} Spacing between nodes is chosen to assign positions based on the ordering and alignment, and to further minimize edge length.
\item {\bf Render:} The nodes and links are drawn, colored, labeled, and interpolated as appropriate.
\end{enumerate}

All optimization of aesthetics ({\bf R2}) occurs during the ordering, alignment, and placement phase, which is similar to the multi-stage approach taken by Liu et al. \cite{liu2013storyflow}.  They first order and align the storylines using discrete optimization techniques, and then determine the placement of the storylines by adjusting the space between the storylines using a continuous optimization strategy (i.e., quadratic programming).  Our overall approach is similar, but the techniques we use within these stages differ significantly, and where appropriate we highlight these differences.  Below we describe our five steps in detail.

\subsection{Ingestion}
The data ingestion phase is responsible for taking the raw data from the user and transforming it into a sequence of events or induced graphs.  This phase is application specific; interested readers can find a comprehensive discussion of dynamic networks and their applications in Ref~\cite{holme2012temporal}.  Below we describe some of the preprocessing that occurs immediately after ingestion.

\subsubsection{Aggregate Graph}
\begin{figure}
\begin{center}
\includegraphics[width=.85\textwidth]{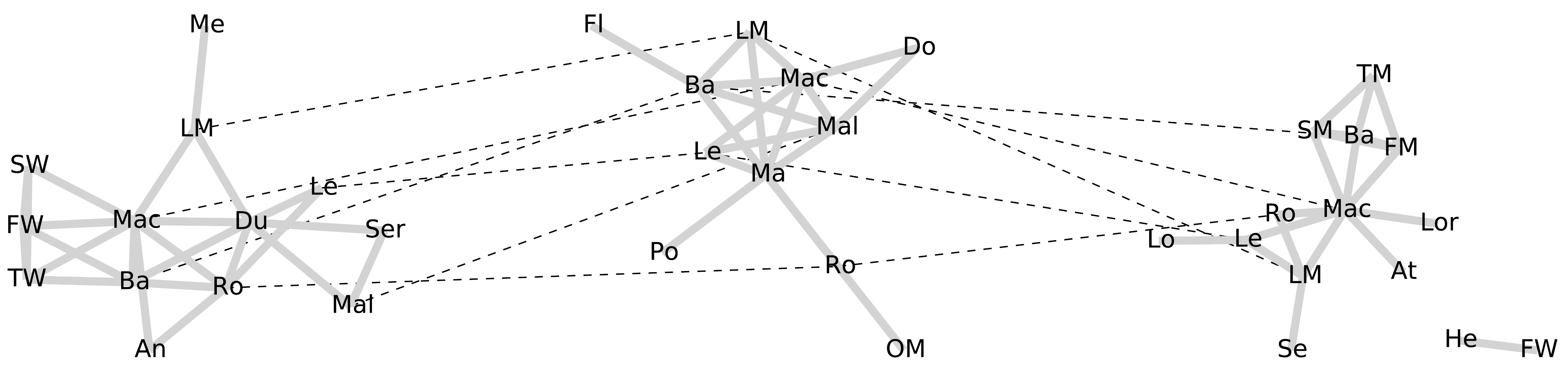}
\caption{The aggregate graph for the first three acts of {\em Macbeth}.  Solid lines indicate edges within the original graphs, and dashed lines indicated edges between the adjacent graphs.}
\end{center}
\label{fig:agg}
\end{figure}

Given a sequence of graphs induced from time windows, an aggregate graph $G_{agg}=(V_{agg},E_{agg})$ is constructed.  If an edge $(u,v)\in E_i$ then $((i,u),(i,v))\in E_{agg}$ with the same weight as the original edge.  Additionally, if $u\in V_i \cap V_{i+1}$ then $((i,u),(i+1,u))\in E_{agg}$.  The weight of this edge can be specified by the user, and larger values will make the corresponding node more likely to be straightened across time windows.  An example of an aggregate graph is shown in Fig.~\ref{fig:agg}.

\subsubsection{Continuation \& Discretization}
Optionally, a more classical view of the storyline can be accomplished via continuation of the storylines.  This is done by calculating $v_{min}$ and $v_{max}$, the first and last appearances of a given node $v$.  For a given time window $i$, if $v_{min} < i < v_{max}$, then $v$ is added to $V_i$ if it is not already present.  The inverse of this operation is discretization, where $v$ is removed from $V_i$ if its degree in $G_i$ is 0.  When necessary, these operations are performed prior to construction of the aggregate graph.

\subsection{Ordering}
The purpose of the ordering phase is to determine an ordering of the nodes within each time window in a manner that supports {\bf R2}.  A good ordering can greatly reduce crossings, of which there are two types: node-node and node-edge.  Node-node crossings occur between time windows when two nodes exist in adjacent scenes and their relative ordering is flipped going from one scene to the next.  Node-edge crossings occur within a time window and are a result of connected nodes not being placed at adjacent levels in the graph.  An example of the two types of crossings is shown in Fig.~\ref{fig:cross}.

The problem of minimizing node-edge crossings and node-node crossings exists in a tradeoff space.  If each node was assigned a unique level, then there would be no node-node crossings, but this would most likely produce a drawing with a large number of node-edge crossings and poor aesthetics.  Similarly, if the node edge crossings were somehow minimized for each time window independently, then we would likely see many node-node crossings.  Therefore, because these two aesthetic properties are in competition with each other, it is the job of the ordering algorithm to balance the two, and provide a means for the user to determine a weighting on their relative importance.  We approach this problem by performing matrix seriation \cite{liiv2010seriation} on the aggregate graph, which induces an ordering of the nodes within each time window.  There are several approaches to the matrix seriation problem, and we consider two of the most popular: spectral seriation and dendrogram seriation.

Our spectral seriation approach is equivalent to a one dimensional ``spectral layout'' of the weighted graph \cite{hagberg-2008-exploring}, and is also very similar to the ``Laplacian eigenmap'' embedding and dimension reduction technique \cite{belkin2001laplacian}.  Essentially, the eigenvector corresponding to the smallest nonzero eigenvalue of the graph laplacian matrix is found.  The nodes are then sorted according to this eigenvector to produce the ordering.  This technique results in a very good approximation of the globally optimal solution, and scalable sparse implementations are readily available.

Liu et al. illustrated the usefulness of placing hierarchical constraints on the ordering of storylines in order to maintain correctness of the drawing.  For this reason we consider an alternative ordering method, dendrogram seriation, which is most commonly known for its use in the ``cluster heat map'' \cite{wilkinson2009history}.  Dendrogram seriation addresses the problem of ordering the leaves in a tree to place similar leaves (according to some arbitrary metric) near to each other.  This ordering must be consistent with the tree structure, and only $2^{n-1}$ of the $n!$ possible permutations of the leaves meet this requirement. There are several algorithms which can solve this problem optimally in polynomial time \cite{bar2001fast,brandes2007optimal}. Then there is the matter of finding the tree in the first place.  The tree may be determined fully or partially from the hierarchical constraints placed on the nodes by the user.  However, it may also be useful to generate the tree using a community detection algorithm, and many algorithms exist for this problem \cite{noack2009multi,fortunato2010community}.  Many of these algorithms are designed to find partitions of the network with high modularity--where clusters of nodes have many intra-cluster edges and few inter-cluster edges.  This property is desirable to help minimize edge crossing ({\bf R2a}) and length ({\bf R2c}), as these nodes will be placed close to each other.

\subsection{Alignment}
\begin{figure}
\begin{center}
\includegraphics[width=.5\textwidth]{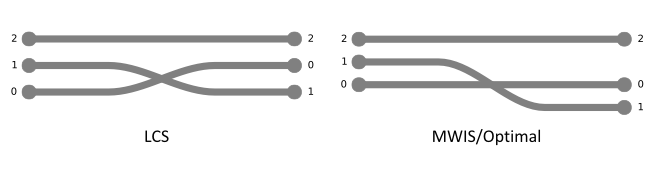}
\end{center}
\caption{Longest common subsequence alignment vs maximum weighted independent set (optimal).  For this example, the longest common subsequence has length 1, and only one node will be aligned regardless of which subsequence is chosen.  A greedy MWIS algorithm will find two nodes to be aligned, which is optimal for this example.}
\label{fig:lcs-counter}
\end{figure}

The purpose of the alignment stage is to improve the readability of the drawing by straightening nodes, or removing ``wiggles.''  The drawing has a wiggle whenever a node is placed at a different height in two adjacent time windows.  In Lin et al.'s alignment phase, wiggles are minimized by solving the longest common subsequence (LCS) problem once per time window, assuming the node order in the previous and current time window are the sequences.  While apparently effective for the datasets presented in the paper, we believe this is not the best approach for the problem.  Fig.~\ref{fig:lcs-counter} demonstrates a typical case where LCS alignment performs poorly.  In general the LCS technique will not scale effectively, as only one (potentially small) subsequence gets aligned, leaving the remaining nodes unaligned.

Instead, we propose solving this problem by framing the alignment problem as a maximum weighted independent set problem (MWIS).  To do so we must first understand that two nodes cannot be simultaneously aligned if those two nodes cross (e.g., nodes 0 and 1 in Fig.~\ref{fig:lcs-counter}), because doing so would change their ordering.  Therefore, we can find all pairs of nodes that cross and construct a constraint graph $G_c$.  For a pair of adjacent time windows, the nodes of the constraint graph are those nodes that are present in both time windows (i.e., nodes that will be connected with integrating lines).  Two nodes are connected in the constraint graph if their integrating lines cross, which can be found directly from the ordering of the nodes.  A set of feasible nodes to align is a subgraph of $G_c$ that has no edges.  Finding the largest such subgraph will align the most nodes, which is precisely the maximal independent set problem.  If the nodes are weighted according to importance by the user (e.g., how important it is for a particular node to be aligned), the problem becomes the MWIS problem where the objective is to find a subgraph of $G_c$ that contains no edges and whose node weight sum is the largest.  While being an NP-hard problem and the subject of much research, several simple greedy heuristics suffice as a fast approximations with good provable bounds \cite{sakai2003note}.

\subsection{Placement}
In the placement phase the vertical positions of each node are determined, subject to the ordering and alignment constraints found in the previous stages of the layout algorithm.  In this stage, the empty space between nodes should be effectively used to shorten edge lengths as much as possible, supporting {\bf R2c}.  Similarly, Liu et al. discuss a ``compaction'' phase, whose objective is to minimize wiggles and unnecessary whitespace.  This is the continuous portion of their hybrid layout algorithm, and they rely on a quadratic program solver to perform the optimization.  Instead, we are able to phrase our placement problem in a graphical, rather than numerical framework, allowing the optimal result to be quickly found.  Given the ordering and alignment constraints we first construct a weighted directed acyclic graph (DAG).  Then, we use the network simplex algorithm \cite{gansner1993technique} to compute an optimal level assignment for the DAG.  The levels found are used directly as the heights of the nodes in the final SVEN layout.  We found the naive $O(|V|^2\cdot |E|)$ version of the network simplex algorithm was straightforward to implement and ran quick enough to be effective for the datasets we examined.  Next we provide some more details on how to construct the DAG described above.

\begin{figure}
\begin{center}
\includegraphics[width=.5\textwidth]{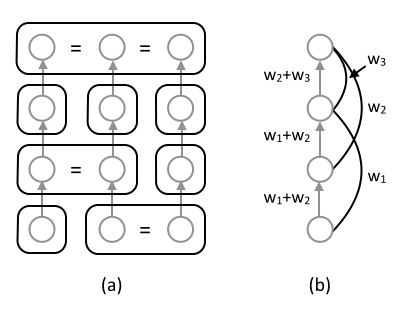}
\end{center}
\caption{Illustration of how to construct a weighted DAG representing the ordering and alignment constraints. (a) The connected component classes of the aligned nodes are connected with directed edges when adjacent. The small circles represent nodes within each time window, and the larger rounded boxes represent the connected component classes resulting from the alignment. (b) Each edge in $E_i$ from time window $i$ contributes its weight to each edge in the DAG it spans.}
\label{fig:dag}
\end{figure}

The first step is to determine the groups of nodes that are being aligned.  This can be accomplished by constructing a graph containing an edge $((i,v),(i+1,v))$ if node $v$ was aligned between time window $i$ and $i+1$.  We denote $C(i,v)$ to be the connected component class of node $v$ in time window $i$ and $s(i,j)$ as the $j^{th}$ node in time window $i$.  The DAG, is constructed by adding edges between connected component classes that enforce the ordering previously found for each time window, and these edges have the form $(C(i,s(i,j)),C(i,s(i,j+1)))$.  An edge $(u,v)$ in $G_i$ contributes a weight of $w(u,v)$ to all edges in the DAG of the form $(C(i,s(i,j)),C(i,s(i,j+1)))$ where $s(i,u) \leq s(i,j) < s(i,v)$.  For clarity, Fig.~\ref{fig:dag} illustrates how to construct the DAG containing the ordering and alignment constraints.

\subsection{Rendering}
We render the storylines using a metro map design, wherever possible drawing straight lines with smooth bends and nameable colors.  When a node appears in consecutive time windows, their storyline is interpolated between the time windows. A node's storyline disappears when that node is not present in the following time window.  The ends of the storylines are capped with a arrows if node's storyline reappears later (or previously), otherwise they are capped with circles.  There are likely to be more nodes than nameable colors, so an application specific means for assigning colors is needed.  By default, we assign nameable colors to the most frequently occurring nodes in the graph, and a light gray color to the remaining nodes.

\section{Results}
Here we briefly show several examples of layouts generated using SVEN from a variety of datasets.

\begin{figure}[htbp]
\begin{center}
\subfigure[\label{fig:ncaaf-noopt}]{\includegraphics[width=0.85\textwidth]{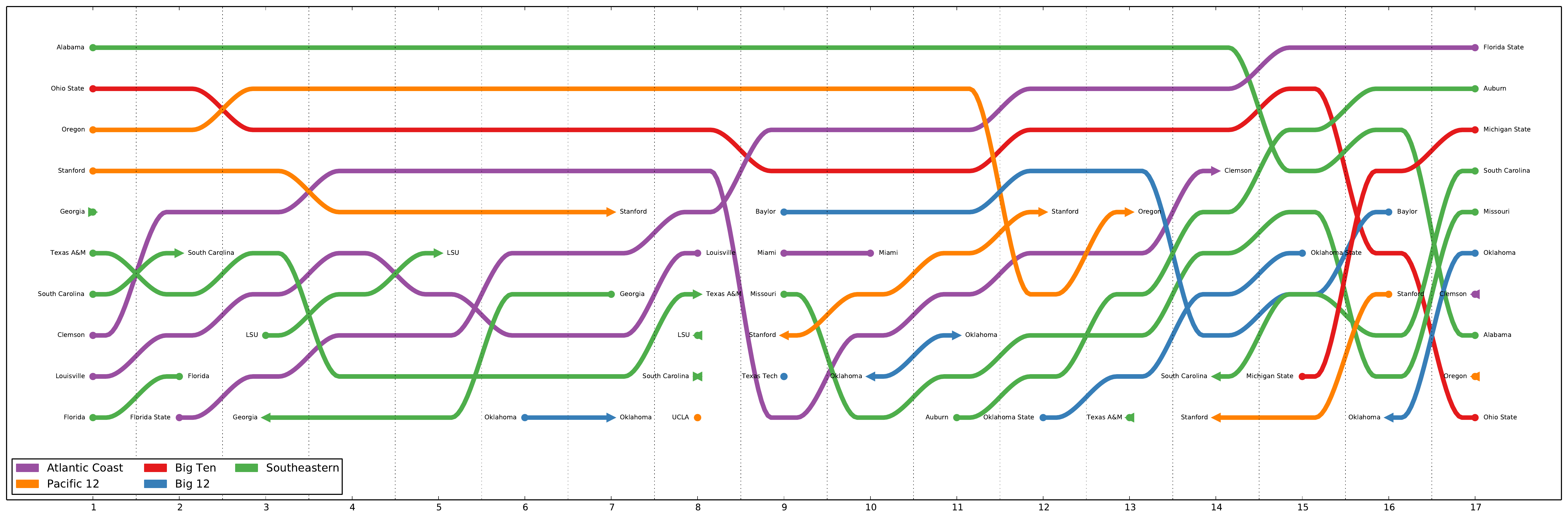}}
\subfigure[\label{fig:ncaaf-best}]{\includegraphics[width=0.85\textwidth]{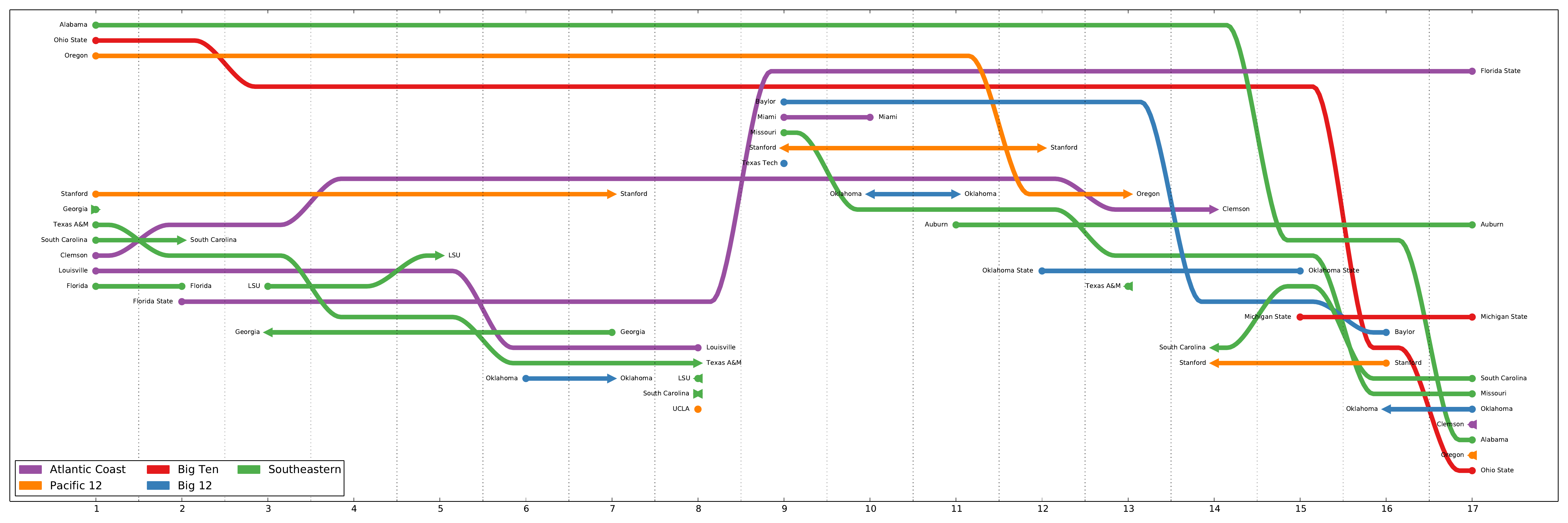}}
\subfigure[\label{fig:ncaaf-truthy}]{\includegraphics[width=0.85\textwidth]{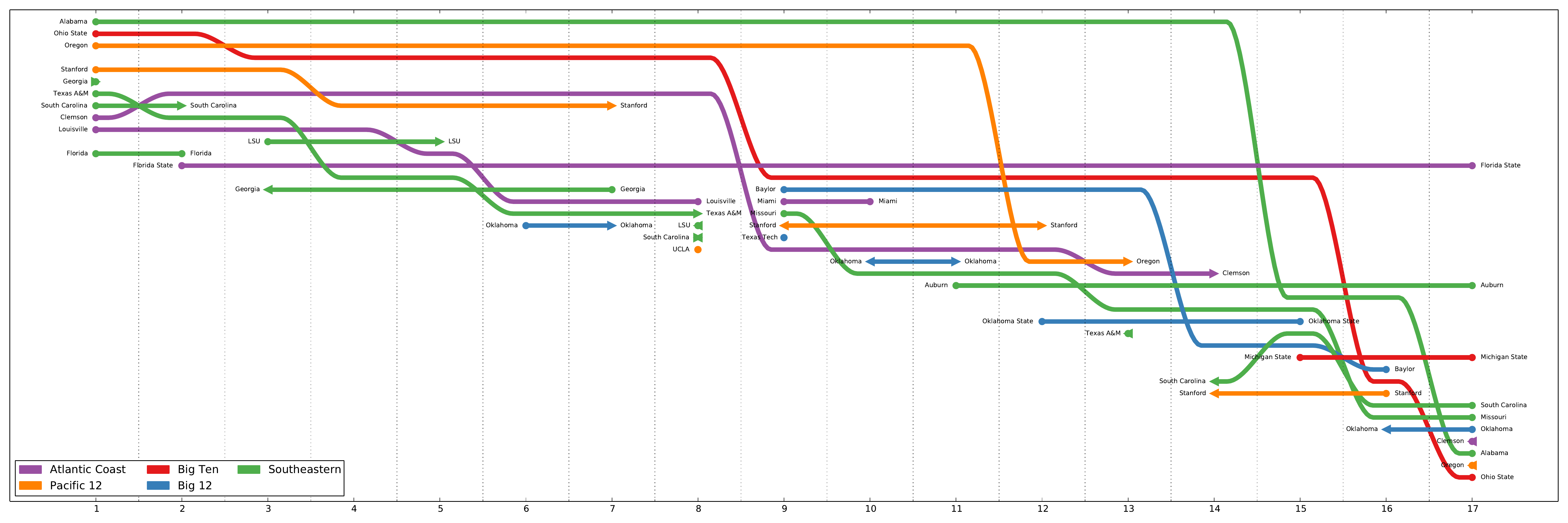}}
\caption{Top 10 NCAAF teams, ranked by the USA Today poll, throughout the 2013 season. (a) Direct mapping of the rankings onto the vertical axis (b) Rankings with adjusted spacing using SVEN's alignment and placement stages. (c) Weighting in the MWIS algorithm chosen to straighten rank increases.}
\label{fig:ncaaf}
\end{center}
\end{figure}

Fig.~\ref{fig:ncaaf} demonstrates the utility of SVEN's alignment and placement stages.  It shows changes in the weekly NCAA Football rankings from the USA Today Poll\footnote{source: ESPN \url{http://espn.go.com/college-football/rankings}} throughout the 2013 season.  Only the top 10 teams from each week are shown.  Fig.~\ref{fig:ncaaf-noopt} shows a direct mapping of the changing rankings using SVEN's metro map rendering style without any optimization (i.e., there is 1 unit of space between each team).  Fig.~\ref{fig:ncaaf-best}, demonstrates the use of SVEN's alignment and placement stages, while maintaining the ordering from the ranking.  We noticed that in this dataset, generally rank increases in the top ranked teams are small whereas rank decreases are large.  A rank increase for some team $A$ will generally occur when another higher ranked team $B$ loses a game, causing $B$ to drop below $A$.  Therefore, in MWIS algorithm nodes are weighted to prefer aligning nodes with rank increases over time (i.e., weights provided to the MWIS algorithm are $1$ unless a team's rank decreases, in which case the weight is $0$), which is shown in Fig.~\ref{fig:ncaaf-truthy}.  The effect of this weighting is a representation of the data that highlights sudden decreases in a team's rank. We note that the representation in Fig.~\ref{fig:ncaaf-truthy} is not a linear encoding of the ranks, which makes reading the true rank of the team directly from the diagram more difficult, but relative rankings are more readily apparent.

\begin{figure}[htbp]
\begin{center}
\includegraphics[width=.85\textwidth]{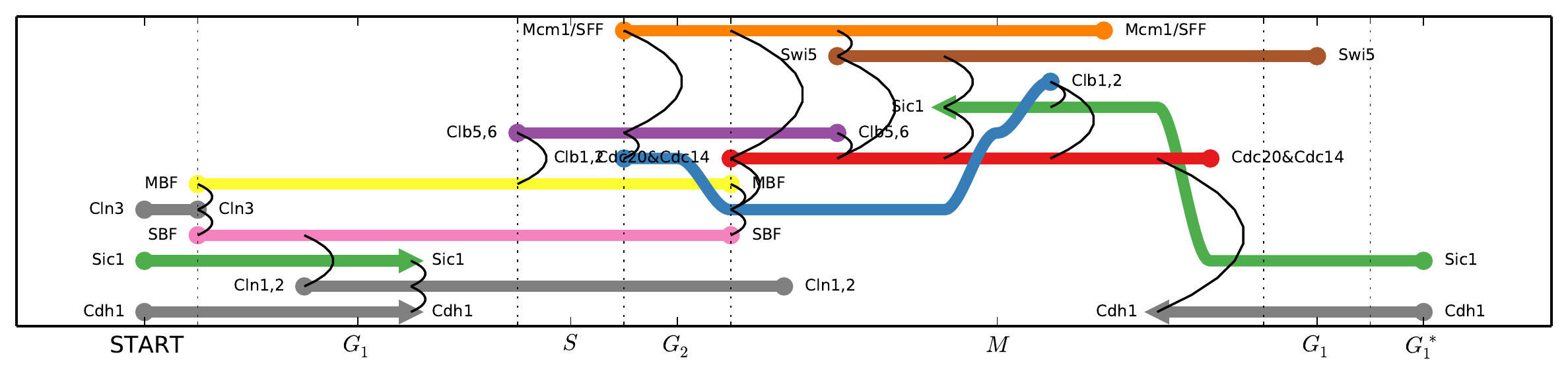}
\caption{Budding yeast cell cycle modeled by \cite{Li2004}.  Edges represent activation or inhibition, which occurs between discrete time steps in the model.}
\label{fig:yeast}
\end{center}
\end{figure}

Fig.~\ref{fig:yeast} is a representation of the dynamics occurring during cell division for a simplified model of budding yeast \cite{Li2004}.  During this process, various proteins interact by activating or inhibiting other proteins, or naturally decay via degradation.  This carefully choreographed sequence of events is responsible for regulating each phase of the cell division.  The process can involve hundreds of different proteins, but the dynamical model developed by Li et al.~\cite{Li2004} greatly summarizes the dynamics in terms of just a handful of key players.  The result of their analysis is a discrete-time discrete-state description of the cell-cycle dynamics.  Combined with the protein interaction network, we can deduce which proteins were responsible for activation and inhibition between each discrete time step, and we represent these as $(t,(u,v))$ events, to be read as ``at time $t$ protein $u$ contributed to the activation or inhibition of protein $v$.''  Note that more than one protein can contribute (i.e., $v$ can have an in-degree that is greater than 1).  We also ensure that if protein $u$ is denoted as active by the dynamical system model at time step $i$, then node $u$ is present in $G_i$ in the visualization.  Additionally, because the dynamics represent a cycle, where the final state and starting states are consistent, we ensure that this is accounted for in the ordering, and alignment stages in SVEN.  The effect of this is seen in Fig.~\ref{fig:yeast} where Sic1 and Cdh1 are present at the same level within the first and last time window.

\begin{figure}[htbp]
\begin{center}
\includegraphics[width=.85\textwidth]{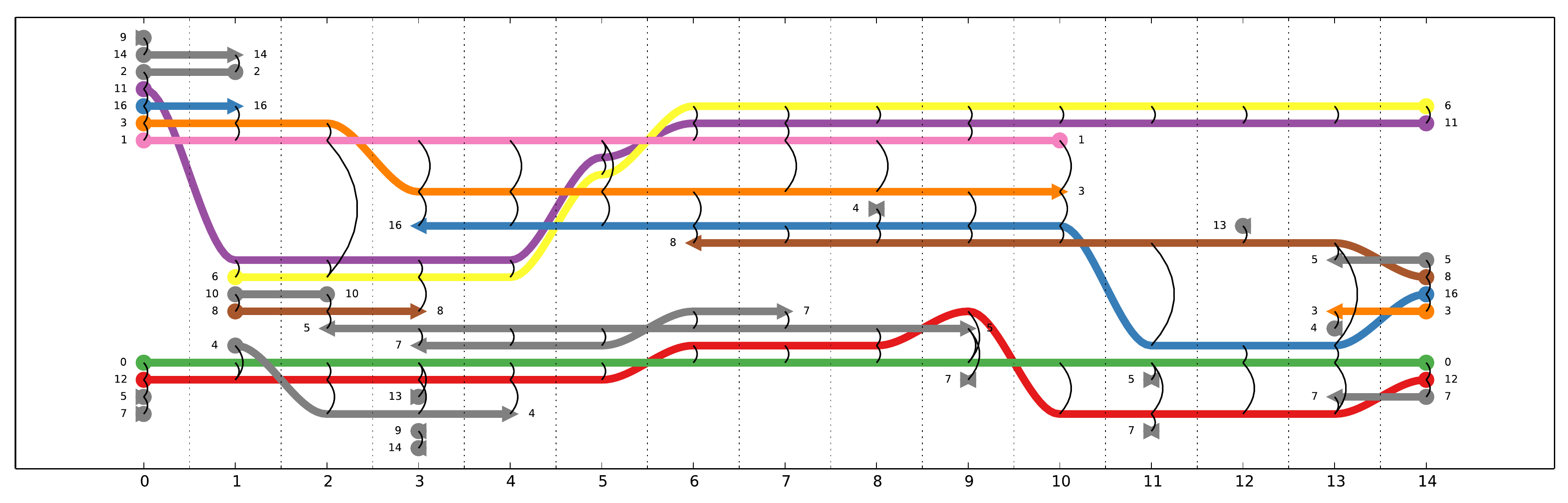}
\caption{The Newcomb Fraternity longitudinal study~\cite{nordlie1958longitudinal,newcomb1961acquaintance,boorman1976social}.  The sequence of graphs is induced by considering close reciprocated friendships.}
\label{fig:frat}
\end{center}
\end{figure}

The ``Newcomb Fraternity'' dataset is a classical longitudinal study of social dynamics~\cite{nordlie1958longitudinal,newcomb1961acquaintance,boorman1976social}.  It contains fifteen $17\times 17$ subjective ranking matrices that represent seventeen individuals' rankings of the other 16 fraternity members, where a low ranking indicates that member views the other member as a close friend.  The fifteen versions of the ranking matrices represent measurement of these rankings over time.  This dataset is transformed into a sequence of graphs as follows.  Let $r_{ij}^t$ be the subjective ranking of individual $j$ by individual $i$ at time $t$.  Then $E_t$ contains the edge $(i,j)$ if and only if $r_{ij}^t \leq \epsilon$ and $r_{ji}^t \leq \epsilon$, where $\epsilon$ is an arbitrarily chosen threshold.  This results in an induced graph for each time window that represents close and reciprocated friendships (we let $\epsilon = 2$, as this is the smallest $\epsilon$ that allows for non-trivial  structures).  Fig.~\ref{fig:frat} shows the SVEN visualization of this dataset.

\begin{figure}[htbp]
\begin{center}
\subfigure[]{\includegraphics[width=.48\textwidth]{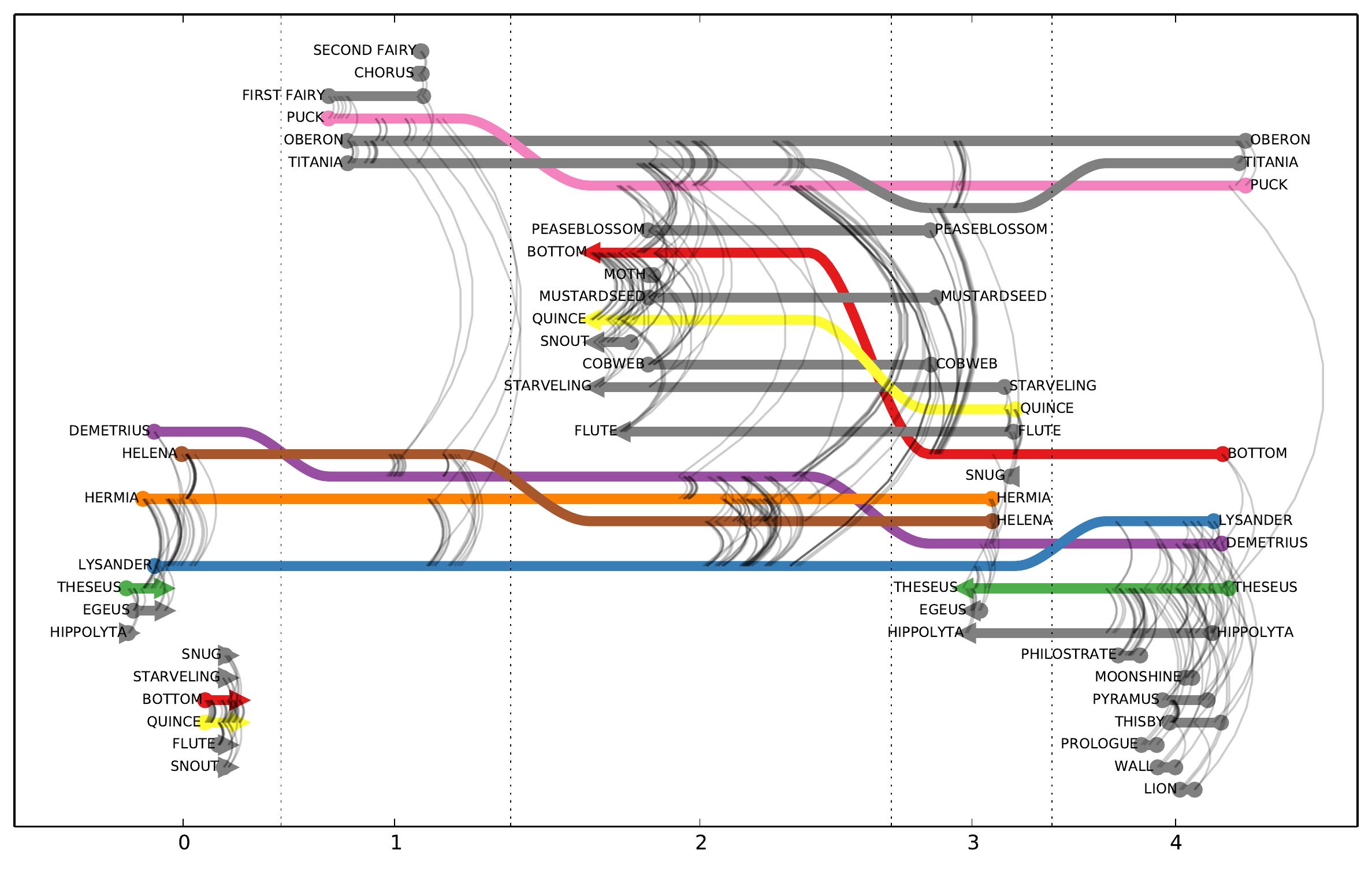}}
\subfigure[]{\includegraphics[width=.48\textwidth]{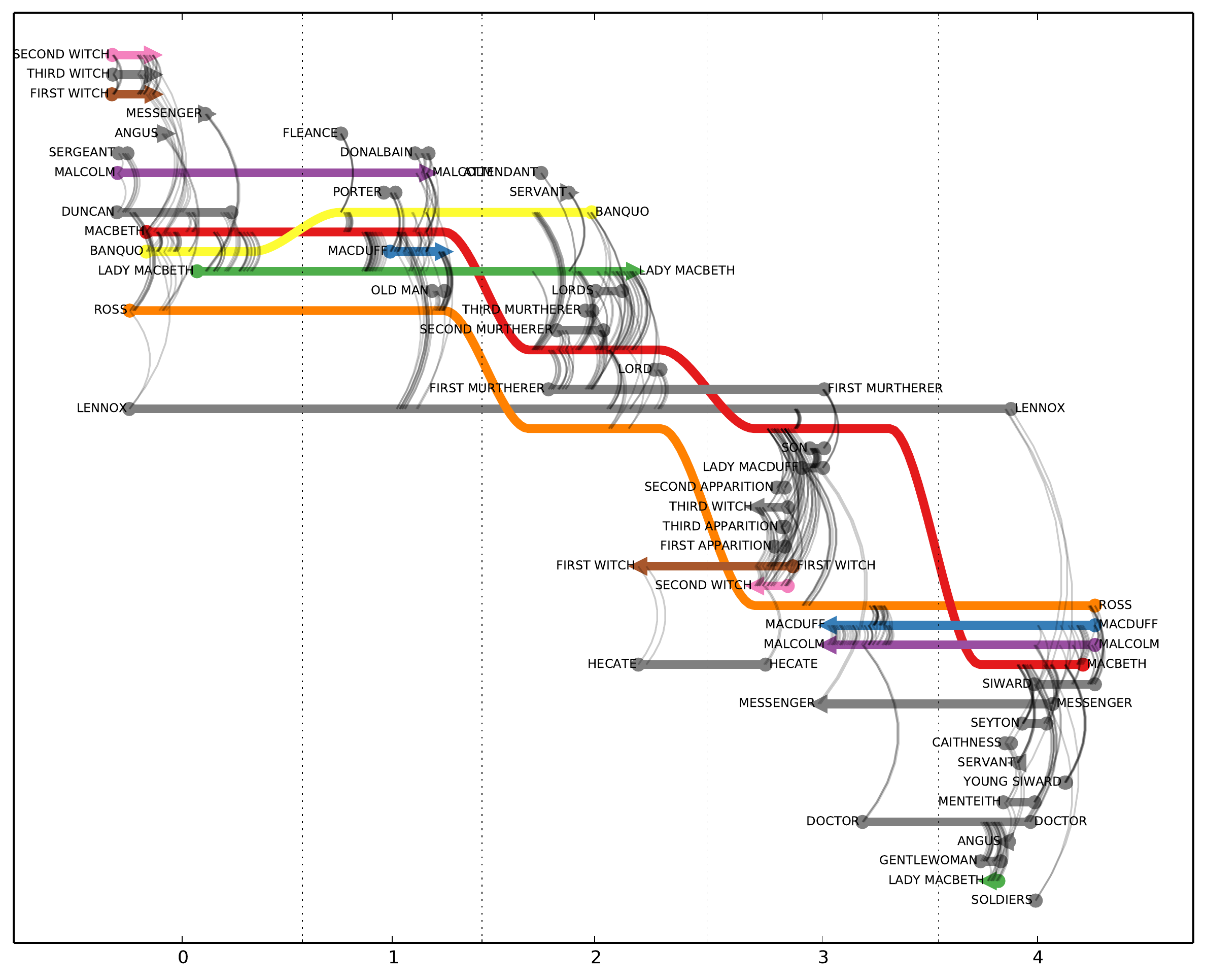}}
\caption{Dialogue from Shakespeare plays. (a) {\em Midsummer Night's Dream} (b) {\em Macbeth}}
\label{fig:plays}
\end{center}
\end{figure}

Dynamic network visualizations can be useful to understand the evolution of discourse (i.e., a sequence of speakers) to understand who talks to whom, and how these relationships change over time.  To process dialogue into a dynamic network, first a parser must extract the sequence of speakers $\{s_1, s_2, ...\}$ from the raw text data.  Then, this sequence is transformed into a set of network events of the form $(t,(s_i,s_{i+1}))$ where $t$ is the line number and $s_i$ and $s_{i+1}$ are consecutive speakers.  The event sequence is then partitioned into time windows; a good discretization of time would be to follow the act or scene breaks (in the case of a play) or chapters (in the case of a book), as these are the natural temporal breaks decided upon by the author.  Fig.~\ref{fig:plays} shows SVEN visualizations of the dialogue in two well known Shakespeare plays\footnote{source: Project Gutenberg  \url{www.gutenberg.org}}.

\section{Conclusions \& Future Work}
We have presented a preliminary algorithm for adapting storyline visualization techniques to the problem of dynamic graph visualization.  Our solution is derived from two basic requirements: representing time using the horizontal axis ({\bf R1}) and reducing clutter by rearranging the storylines along the vertical axis ({\bf R2}).  We approach this challenging problem by breaking the optimization into multiple stages (ordering, alignment, and placement) similar to previous work \cite{liu2013storyflow}.  However, we differ significantly from previous work in how we accomplish these sub-tasks.  Ordering of nodes is quickly and effectively accomplished by seriation of the aggregate graph matrix using spectral methods (alternatively, hierarchical graph clustering and dendrogram seriation can be used).  Alignment of the nodes is performed by determining node crossings between adjacent time windows, and solving the maximum weighted independent set (MWIS) problem to maximize the number of straightened nodes.  Then the placement of the nodes is optimized using the network simplex algorithm.

We have applied SVEN to a variety of datasets.  We showed how SVEN can be used to optimize the placement of rankings over time and that the weighting of nodes for the MWIS algorithm can have a beneficial effect on the readability of the diagram when properly chosen.  We utilized SVEN to visualize the complex dynamics produced by a model of the budding yeast cell cycle, clarifying the sequence of activation, inhibition, and degradation that regulates the process.  SVEN was used to visualize a classical longitudinal study of social acquaintances in a fraternity, highlighting close friendships and changing groups.  Finally, SVEN was used to show the sequence of speakers in two well known Shakespeare plays.  Much future work remains, and this work falls into three categories: implementing interactivity, performing evaluation, and addressing issues pertaining to scalability.

\subsection{Interactivity}
Interactivity is crucial to accommodate some of the shortcomings inherent to SVEN visualizations of dynamic networks.  Many potentially useful interactions fall into the category of ``details on demand,'' where additional information can be given to the user when they request it.  For example, when a node is absent from a time window, but present before and after, that node's storyline will be broken making the line harder to trace through the visualization.  A solution could be to complete the storyline for a given node at the user's request.  Another useful way to provide details is to allow the user to brush across the visualization, selecting a subset of the nodes and time simultaneously.  Then an alternative visualization, such as a node link diagram or adjacency matrix, can be shown to the user for this particular subset where it may be more informative than the overview provided by SVEN.

Temporal networks can have a large number of edges which can cause clutter.  A solution would be to use transparency or other techniques to de-emphasize these edges, but then user may want to re-emphasize all the edges incident to a particular node over time or at a particular time upon request. Other interactions could involve supporting various analyses relevant to temporal networks, such as temporal shortest path.  For example, the user could select a node at a particular time and request that the visualization shows when other nodes can be reached, the minimum number of steps this requires, and the particular optimal path(s) to accomplish this.  Additional application specific details should also be presented directly in the visualization, or made available through interaction.  Other standard interactions such as adding/deleting nodes/links and re-arranging the visualization would be important features of a fully functioning visual analytics tool for dynamic network visualization.

\subsection{Evaluation}
Evaluation of SVEN is an important future research activity--many unanswered questions remain.  Crucially, we would like empirical evidence about what combinations of data and task allow users of SVEN to attain good performance relative to existing techniques.  This can be accomplished with a controlled user study with simple tasks that have measurable performance.  In this case it may be important to generate the datasets from a parametrized random model of a temporal networks, rather than using real-world datasets.  To get a better understanding of how SVEN could provide insight to an analyst, the opposite is the case--datasets should be taken from the real world with a particular application and challenging question in mind.  Evaluation by the ``analysts'' would be more subjective compared to the controlled user study, but potentially more informative.  Additionally, objective evaluation of the layout algorithm against previous work, different node/link weightings, or different heuristics can be accomplished by measuring the aesthetic qualities of the resulting layouts (e.g., crossings, wiggles, line length).

\subsection{Scalability}
From a computational standpoint, SVEN is scalable to large datasets\footnote{In the analysis of SVEN's time complexity, we assume the number of time windows is a small constant, which is bounded more by the available screen space than the data.}.  The ordering stage's time complexity, when using the spectral layout method, which requires computing the eigenvectors of a matrix, is $O(|V|^3)$ or better depending on sparsity.  The alignment stage requires determining the set of nodes that cross between adjacent time windows.  Our naive implementation simply compares all pairs of nodes, which has a time complexity of $O(|V|^2)$; a more sophisticated algorithm using bisection would have a time complexity of $O(|V|\log |V|)$ assuming the number of crossings per node is bounded by a small constant.  Following this, the MWIS algorithm is run; our naive implementation of this is $O(|V|\cdot |E|)$, but can be improved by using a heap implementation of a priority queue with a increase/decrease key functionality.  The placement stage requires running the network simplex algorithm; our naive implementation has a time complexity of $O(|V|^2 \cdot |E|)$, making this the most computationally intensive component of SVEN.  However, even with an overall cubic worst case complexity, the layout can be computed under interactive conditions (i.e., latency $< 500$ms) for thousands of nodes and edges.  We would expect the visualization to become cluttered and unusable long before the layout algorithm starts to running too slowly from the user's perspective.  Therefore, we suggest that future work should prioritize addressing the problem of scalability through effective interaction and abstraction techniques over fine tuning of the layout algorithm.

Common ways of addressing scalability issues are through filtering and abstraction.  For example, filtering can be used to reduce the number of nodes and edges to be shown, which in turn makes the layout easier to compute and potentially less cluttered.  Filtering can be automatic by using degree of interest (DOI) techniques to assign a score to data elements, and visualizing a relevant subset of this data \cite{van2009search}.  For networks, abstraction is frequently accomplished through clustering \cite{fortunato2010community,noack2004energy,abello2006ask,henry2007nodetrix,balzer2007level}, where groups of nodes are aggregated together, and represented as a super node.  In the visualization, more emphasis is placed on the clusters and the inter-cluster links, because clusters are assumed to be dense, and intra-cluster links are less informative.  This technique can be adapted to dynamic networks by leveraging dynamic network clustering techniques.  A simple approach is to compute a hierarchical clustering directly on the aggregate graph, and to use dendrogram seriation to compute the ordering of the nodes, allowing groups to be drawn directly on the visualization in a consistent manner.

\section*{Acknowledgements}
The authors would like to thank Paul Havig at AFRL for his guidance and helpful comments throughout the project.  This research is supported by AFOSR LRIR 12RH12COR to L.M.B., and was performed while D.L.A. held a National Research Council Research Associateship Award at the the Air Force Research Laboratory, Wright-Patterson AFB, Ohio.

\bibliographystyle{plain}
\bibliography{refs}

\begin{thebibliography}{10}

\bibitem{abello2006ask}
James Abello, Frank Van~Ham, and Neeraj Krishnan.
\newblock Ask-graphview: A large scale graph visualization system.
\newblock {\em Visualization and Computer Graphics, IEEE Transactions on},
  12(5):669--676, 2006.

\bibitem{aigner2011timebook}
Wolfgang Aigner, Silvia Miksch, Heidrum Schumann, and Christian Tominski.
\newblock {\em Visualization of Time-Oriented Data}.
\newblock Springer, London, 2011.

\bibitem{archambault2011}
Daniel Archambault, Helen Purchase, and Bruno Pinaud.
\newblock Animation, small multiples, and the effect of mental map preservation
  in dynamic graphs.
\newblock {\em Visualization and Computer Graphics, {IEEE} Transactions on},
  17(4):539--552, 2011.

\bibitem{balzer2007level}
Michael Balzer and Oliver Deussen.
\newblock Level-of-detail visualization of clustered graph layouts.
\newblock In {\em Visualization, 2007. APVIS'07. 2007 6th International
  Asia-Pacific Symposium on}, pages 133--140. IEEE, 2007.

\bibitem{bar2001fast}
Ziv Bar-Joseph, David~K Gifford, and Tommi~S Jaakkola.
\newblock Fast optimal leaf ordering for hierarchical clustering.
\newblock {\em Bioinformatics}, 17(suppl 1):S22--S29, 2001.

\bibitem{bastian2009}
Mathieu Bastian, Sebastien Heymann, and Mathieu Jacomy.
\newblock Gephi: an open source software for exploring and manipulating
  networks.
\newblock In {\em ICWSM}, 2009.

\bibitem{belkin2001laplacian}
Mikhail Belkin and Partha Niyogi.
\newblock Laplacian eigenmaps and spectral techniques for embedding and
  clustering.
\newblock In {\em NIPS}, volume~14, pages 585--591, 2001.

\bibitem{bender2006art}
Skye Bender-{deMoll} and Daniel~A {McFarland}.
\newblock The art and science of dynamic network visualization.
\newblock {\em Journal of Social Structure}, 7(2):1--38, 2006.

\bibitem{bendix2005parallel}
Fabian Bendix, Robert Kosara, and Helwig Hauser.
\newblock Parallel sets: Visual analysis of categorical data.
\newblock In {\em Information Visualization, 2005. INFOVIS 2005. IEEE Symposium
  on}, pages 133--140. IEEE, 2005.

\bibitem{berger2006framework}
Tanya~Y Berger-Wolf and Jared Saia.
\newblock A framework for analysis of dynamic social networks.
\newblock In {\em Proceedings of the 12th ACM SIGKDD international conference
  on Knowledge discovery and data mining}, pages 523--528. ACM, 2006.

\bibitem{boorman1976social}
Scott~A Boorman and Harrison~C White.
\newblock Social structure from multiple networks. ii. role structures.
\newblock {\em American Journal of Sociology}, pages 1384--1446, 1976.

\bibitem{bostock2011}
Michael Bostock, Vadim Ogievetsky, and Jeffrey Heer.
\newblock D$^3$ data-driven documents.
\newblock {\em IEEE Transactions on Visualization and Computer Graphics},
  17(12):2301--2309, 2011.

\bibitem{brandes2007optimal}
Ulrik Brandes.
\newblock Optimal leaf ordering of complete binary trees.
\newblock {\em Journal of Discrete Algorithms}, 5(3):546--552, 2007.

\bibitem{brandes2003}
Ulrik Brandes and Steven~R Corman.
\newblock Visual unrolling of network evolution and the analysis of dynamic
  discourse.
\newblock {\em Information Visualization}, 2(1):40--50, 2003.

\bibitem{burch2011parallel}
Michael Burch, Corinna Vehlow, Fabian Beck, Stephan Diehl, and Daniel Weiskopf.
\newblock Parallel edge splatting for scalable dynamic graph visualization.
\newblock {\em Visualization and Computer Graphics, {IEEE} Transactions on},
  17(12):2344--2353, 2011.

\bibitem{chen2006}
Chaomei Chen.
\newblock {CiteSpace-II}: {D}etecting and visualizing emerging trends and
  transient patterns in scientific literature.
\newblock {\em Journal of the American Society for Information Science and
  Technology}, 57(3):359--377, 2006.

\bibitem{dwyer2009}
Tim Dwyer.
\newblock Scalable, versatile and simple constrained graph layout.
\newblock {\em Computer Graphics Forum}, 28(3):991--998, 2009.

\bibitem{dwyer2002visualising}
Tim Dwyer and Peter Eades.
\newblock Visualising a fund manager flow graph with columns and worms.
\newblock In {\em Information Visualisation, 2002. Proceedings. Sixth
  International Conference on}, pages 147--152. IEEE, 2002.

\bibitem{ellson2002}
John Ellson, Emden Gansner, Lefteris Koutsofios, Stephen~C North, and Gordon
  Woodhull.
\newblock Graphviz--open source graph drawing tools.
\newblock {\em Graph Drawing}, pages 483---484, 2002.

\bibitem{erten2004}
Cesim Erten, Philip~J Harding, Stephen~G Kobourov, Kevin Wampler, and Gary Yee.
\newblock Exploring the computing literature using temporal graph
  visualization.
\newblock In {\em Electronic Imaging 2004}, pages 45--56. International Society
  for Optics and Photonics, 2004.

\bibitem{fortunato2010community}
Santo Fortunato.
\newblock Community detection in graphs.
\newblock {\em Physics Reports}, 486(3):75--174, 20120.

\bibitem{fruchterman1991drawing}
Thomas~M Fruchterman and Edward~M Reingold.
\newblock Graph drawing by force-directed placement.
\newblock {\em Software--Practice and Experience}, 21(1):1129--1164, 1991.

\bibitem{gansner1993technique}
Emden~R Gansner, Eleftherios Koutsofios, Stephen~C North, and K-P Vo.
\newblock A technique for drawing directed graphs.
\newblock {\em Software Engineering, IEEE Transactions on}, 19(3):214--230,
  1993.

\bibitem{hagberg-2008-exploring}
Aric~A. Hagberg, Daniel~A. Schult, and Pieter~J. Swart.
\newblock Exploring network structure, dynamics, and function using {NetworkX}.
\newblock In {\em Proceedings of the 7th Python in Science Conference
  (SciPy2008)}, pages 11--15, Pasadena, CA USA, August 2008.

\bibitem{henry2007nodetrix}
Nathalie Henry, J~Fekete, and Michael~J McGuffin.
\newblock Nodetrix: a hybrid visualization of social networks.
\newblock {\em Visualization and Computer Graphics, IEEE Transactions on},
  13(6):1302--1309, 2007.

\bibitem{holme2012temporal}
Petter Holme and Jari Saram{\"a}ki.
\newblock Temporal networks.
\newblock {\em Physics reports}, 519(3):97--125, 2012.

\bibitem{chi1999}
Ed~Huai hsin Chi and Stuart~K Card.
\newblock Sensemaking of evolving web sites using visualization spreadsheets.
\newblock In {\em Information Visualization, 1999 (Info VIs' 99) Proceedings.
  1999 {IEEE} Symposium on}, pages 18--25, 1999.

\bibitem{javed2012exploring}
Waqas Javed and Niklas Elmqvist.
\newblock Exploring the design space of composite visualization.
\newblock In {\em Pacific Visualization Symposium (PacificVis), 2012 IEEE},
  pages 1--8. IEEE, 2012.

\bibitem{kamada}
Tomihisa Kamada and Satoru Kawai.
\newblock An algorithm for drawing general undirected graphs.
\newblock {\em Information Processing Letters}, 31:7--15, 1989.

\bibitem{kim2010tracing}
Nam~Wook Kim, Stuart~K Card, and Jeffrey Heer.
\newblock Tracing genealogical data with timenets.
\newblock In {\em Proceedings of the International Conference on Advanced
  Visual Interfaces}, pages 241--248. ACM, 2010.

\bibitem{lamport1978time}
Leslie Lamport.
\newblock Time, clocks, and the ordering of events in a distributed system.
\newblock {\em Communications of the ACM}, 21(7):558--565, 1978.

\bibitem{Li2004}
Fangting Li, Tao Long, Ying Lu, Qi~Ouyang, and Chao Tang.
\newblock {The yeast cell-cycle network is robustly designed}.
\newblock {\em Proceedings of the National Academy of Sciences of the United
  States of America}, 101(14):4781--4786, 2004.

\bibitem{liiv2010seriation}
Innar Liiv.
\newblock Seriation and matrix reordering methods: An historical overview.
\newblock {\em Statistical analysis and data mining}, 3(2):70--91, 2010.

\bibitem{liu2013storyflow}
Shixia Liu, Yingcai Wu, Enxun Wei, Mengchen Liu, and Yang Liu.
\newblock Storyflow: Tracking the evolution of stories.
\newblock {\em Visualization and Computer Graphics, IEEE Transactions on},
  19(12):2436--2445, 2013.

\bibitem{moody2005dynamic}
J.~Moody, D.~McFarland, and S.~Bender-deMoll.
\newblock Dynamic network visualization.
\newblock {\em American Journal of Sociology}, 110(4):1206--1241, 2005.

\bibitem{xkcd}
R~Munroe.
\newblock \#657: Movie narrative charts.

\bibitem{newcomb1961acquaintance}
Theodore~M Newcomb.
\newblock The acquaintance process.
\newblock 1961.

\bibitem{noack2004energy}
Andreas Noack.
\newblock An energy model for visual graph clustering.
\newblock In {\em Graph Drawing}, pages 425--436. Springer, 2004.

\bibitem{noack2009multi}
Andreas Noack and Randolf Rotta.
\newblock Multi-level algorithms for modularity clustering.
\newblock In {\em Experimental Algorithms}, pages 257--268. Springer, 2009.

\bibitem{nordlie1958longitudinal}
Peter~Gerhard Nordlie.
\newblock {\em A longitudinal study of interpersonal attraction in a natural
  group setting.}
\newblock PhD thesis, 1958.

\bibitem{ogawa2010software}
Michael Ogawa and Kwan-Liu Ma.
\newblock Software evolution storylines.
\newblock In {\em Proceedings of the 5th international symposium on Software
  visualization}, pages 35--42. ACM, 2010.

\bibitem{purchase2007important}
H.~Purchase, E.~Hoggan, and C.~G{\"o}rg.
\newblock How important is the ``mental map''?--an empirical investigation of a
  dynamic graph layout algorithm.
\newblock In {\em Graph drawing}, pages 184--195. Springer, 2007.

\bibitem{reda2011visualizing}
Khairi Reda, Chayant Tantipathananandh, Andrew Johnson, Jason Leigh, and Tanya
  Berger-Wolf.
\newblock Visualizing the evolution of community structures in dynamic social
  networks.
\newblock In {\em Computer Graphics Forum}, volume~30, pages 1061--1070. Wiley
  Online Library, 2011.

\bibitem{rosvall2010mapping}
Martin Rosvall and Carl~T Bergstrom.
\newblock Mapping change in large networks.
\newblock {\em PloS one}, 5(1):e8694, 2010.

\bibitem{sakai2003note}
Shuichi Sakai, Mitsunori Togasaki, and Koichi Yamazaki.
\newblock A note on greedy algorithms for the maximum weighted independent set
  problem.
\newblock {\em Discrete Applied Mathematics}, 126(2):313--322, 2003.

\bibitem{schmidt2008sankey}
Mario Schmidt.
\newblock The sankey diagram in energy and material flow management.
\newblock {\em Journal of industrial ecology}, 12(1):82--94, 2008.

\bibitem{shaw2009structure}
Blake Shaw and Tony Jebara.
\newblock Structure preserving embedding.
\newblock In {\em Proceedings of the 26th Annual International Conference on
  Machine Learning}, pages 937--944, 2009.

\bibitem{shi2011dynamic}
Lei Shi, Chen Wang, and Zhen Wen.
\newblock Dynamic network visualization in 1.5 {D}.
\newblock In {\em Pacific Visualization Symposium (PacificVis), 2011 {IEEE}}.
  {IEEE}, 2011.

\bibitem{tamassia1986visibility}
Roberto Tamassia and Ioannis~G Tollis.
\newblock A unified approach to visibility representations of planar graphs.
\newblock {\em Discrete \& Computational Geometry}, 1(1):321--341, 1986.

\bibitem{tanahashi2012design}
Yuzuru Tanahashi and Kwan-Liu Ma.
\newblock Design considerations for optimizing storyline visualizations.
\newblock {\em Visualization and Computer Graphics, IEEE Transactions on},
  18(12):2679--2688, 2012.

\bibitem{tantipathananandh2009constant}
Chayant Tantipathananandh and Tanya Berger-Wolf.
\newblock Constant-factor approximation algorithms for identifying dynamic
  communities.
\newblock In {\em Proceedings of the 15th ACM SIGKDD international conference
  on Knowledge discovery and data mining}, pages 827--836. ACM, 2009.

\bibitem{tantipathananandh2007framework}
Chayant Tantipathananandh, Tanya Berger-Wolf, and David Kempe.
\newblock A framework for community identification in dynamic social networks.
\newblock In {\em Proceedings of the 13th ACM SIGKDD international conference
  on Knowledge discovery and data mining}, pages 717--726. ACM, 2007.

\bibitem{van2009search}
Frank Van~Ham and Adam Perer.
\newblock ``search, show context, expand on demand'': Supporting large graph
  exploration with degree-of-interest.
\newblock {\em Visualization and Computer Graphics, IEEE Transactions on},
  15(6):953--960, 2009.

\bibitem{wasserman1994}
Stanley Wasserman and Katherine Faust.
\newblock {\em Social network analysis: Methods and applications}.
\newblock Cambridge University Press, 1994.

\bibitem{wattenberg2002arc}
Martin Wattenberg.
\newblock Arc diagrams: Visualizing structure in strings.
\newblock In {\em Information Visualization, 2002. INFOVIS 2002. IEEE Symposium
  on}, pages 110--116. IEEE, 2002.

\bibitem{wilkinson2009history}
Leland Wilkinson and Michael Friendly.
\newblock The history of the cluster heat map.
\newblock {\em The American Statistician}, 63(2), 2009.

\end{thebibliography}

\end{document}